\documentclass[letterpaper]{emulateapj}

\usepackage{psfig}
\usepackage{rotating}

\newcommand{\hst}{\textit{HST}}
\newcommand{\hstlong}{\textit{Hubble Space Telescope}}

\newcommand{\Chandra}{{\em Chandra}}
\newcommand{\chandra}{\textit{Chandra}}
\newcommand{\chandralong}{\textit{Chandra X-ray Observatory}}

\newcommand{\nh}{\mbox{$N_{\rm H}$}} 
\newcommand{\nhtt}{\mbox{$N_{\rm H,22}$}}

\newcommand{\kteff}{\mbox{$kT_{\rm eff}$}}
\newcommand{\rinfty}{\mbox{$R_{\infty}$}}
\newcommand{\rns}{\mbox{$R_{\rm NS}$}} 
\newcommand{\mns}{\mbox{$M_{\rm NS}$}}

\newcommand{\chisq}{\mbox{$\chi^2$}}
\newcommand{\Chisq}[3]{$\chi^2_\nu$/dof (prob.) = {#1}/{#2} (#3)}
\newcommand{\chisqrnu}{$\chi^2_\nu$}

\newcommand{\xray}{\mbox{X-ray}}

\newcommand{\simlt}{\mathrel{\hbox{\rlap{\hbox{\lower4pt\hbox{$\sim$}}}\hbox{$<$}}}}
\newcommand{\simgt}{\mathrel{\hbox{\rlap{\hbox{\lower4pt\hbox{$\sim$}}}\hbox{$>$}}}}

\newcommand{\approxlt}{\mbox{$\,^{<}\hspace{-0.24cm}_{\sim}\,$}}
\newcommand{\ee}[1]{\mbox{$10^{#1}$}}
\newcommand{\tee}[1]{\mbox{$\times 10^{#1}$}}
\newcommand{\ud}[2]{\mbox{$^{+ #1}_{- #2}$}}
\newcommand{\ppm}{\mbox{$\pm$}}

\newcommand{\unit}[1]{\mbox{$\rm\,#1$}}
\def\deg{\hbox{$^\circ$}}
\def\arcdeg{\hbox{$^\circ$}}
\def\arcmin{\hbox{$^\prime$}}
\def\arcsec{\hbox{$^{\prime\prime}$}}
\def\sec{\mbox{$\,{\rm s}$}}
\newcommand{\msun}{\mbox{$\,M_\odot$}}

\newcommand{\km}{\hbox{$\,{\rm km}$}}

\newcommand{\MeV}{\mbox{$\,{\rm MeV}$}}
\newcommand{\keV}{\mbox{$\,{\rm keV}$}}
\newcommand{\eV}{\mbox{$\,{\rm eV}$}}
\newcommand{\ksec}{\mbox{$\,{\rm ks}$}}

\newcommand{\kpc}{\mbox{$\,{\rm kpc}$}}

\newcommand{\persec}{\mbox{$\,{\rm s^{-1}}$}}

\newcommand{\percmsq}{\mbox{$\,{\rm cm^{-2}}$}}

\newcommand{\peryear}{\mbox{$\,{\rm yr^{-1}}$}}

\newcommand{\cgsflux}{\mbox{$\,{\rm erg\,\percmsq\,\persec}$}}
\newcommand{\cgslum}{\mbox{$\,{\rm erg\,\persec}$}}
 
\newcommand{\cgsdepth}{\mbox{$\,{\rm g\percmsq}$}} 
\newcommand{\cgsaccel}{\mbox{$\,{\rm cm\,s^{-2}}$}} 

\def\OmCen{\mbox{$\omega$Cen}}

\begin{document}

\title{Neutron Star Radius Measurement with the Quiescent Low-Mass X-ray Binary U24 in NGC~6397}

\author{Sebastien Guillot$^1$, Robert E. Rutledge$^1$, Edward F. Brown$^2$}

 \affil{$^1$ Department of Physics, McGill University, 3600 rue
University, Montreal, QC, H3A-2T8, Canada \\ $^2$Department of
Phys. and Astr., Michigan State University, 3250 Biomed.
Phys. Sci. Building, East Lansing, MI 48824-2320, USA}
\email{guillots@physics.mcgill.ca, rutledge@physics.mcgill.ca}

\slugcomment{Draft re-submitted to ApJ on \today}
\shorttitle{\Chandra\ observations of qLMXBs in NGC 6397}

\begin{abstract}
This paper reports the spectral and timing analyses of the quiescent
low-mass X-ray binary (qLMXB) U24 observed during five archived
\Chandra/ACIS exposures of the nearby globular cluster NGC~6397, for a
total of 350\ksec.  We find that the \xray\ flux and the parameters of
the hydrogen atmosphere spectral model are consistent with those
previously published for this source.  On short timescales, we find no
evidence of aperiodic intensity variability, with 90\% confidence
upper limits during five observations ranging between $<$8.6\%
rms and $<$19\% rms, in the 0.0001--0.1 Hz frequency range
(0.5--8.0 keV); and no evidence of periodic variability, with maximum
observed powers in this frequency range having a chance probability of
occurrence from a Poisson-deviated light curve in excess of 10\%.  We
also report the improved neutron star (NS) physical radius measurement,
with statistical accuracy of the order of $\sim$10\%: $\rns =
8.9\ud{0.9}{0.6}\km$ for $\mns = 1.4\msun$.  Alternatively, we provide
the confidence regions in mass--radius space as well as the best-fit
projected radius $\rinfty = 11.9\ud{1.0}{0.8}\km$, as seen by an
observer at infinity.  The best-fit effective temperature, $\kteff =
80\ud{4}{5}\eV$, is used to estimate the neutron star core temperature
which falls in the range $T_{\rm core} = \left( 3.0 - 9.8 \right)
\tee{7}\unit{K}$, depending on the atmosphere model considered. This
makes U24 the third most precisely measured NS radius among
qLMXBs, after those in \OmCen, and in M13.
\end{abstract}


\maketitle


\section{Introduction}
The emission from low-mass \xray\ binaries in quiescence (qLMXBs) is
routinely studied to provide useful constraints on the physical models
of the interior of neutron stars (NSs).  The low luminosity
($\ee{32}$--$\ee{33}\cgslum$, 4--5 orders of magnitude lower than the
outburst luminosities) of these objects was first observed in the
post-outburst stages of the transient LMXBs Cen~X-4 and Aql~X-1
\citep{vanparadijs87}, and initially interpreted as a thermal
blackbody emission powered by some low-level mass accretion onto the
NS surface \citep{verbunt94}.

In an alternate explanation for the energy source of qLMXBs, the
luminosity is provided by the energy released during accretion
episodes by pressure-sensitive nuclear reactions (electron captures,
neutron emission, and pycnonuclear reactions) in the NS deep crust.
The theory of deep crustal heating (DCH; \citealt{brown98}) describes
how the accreted matter piles up at the top of the NS surface, forcing
the matter underneath to deeper layers of the crust, and how the
energy is deposited in the crust.  The resulting nuclear reaction
chain releases $\sim1.5\MeV$ per accreted nucleon (see \cite{gupta07}
and \cite{haensel08} for details about crustal heating models), and
gives rise to a time-average luminosity directly proportional to the
time-averaged accretion rate:
\begin{equation}
  \langle L\rangle = 9\tee{32}\,\frac{\langle \dot{M}
    \rangle}{10^{-11}\unit{\msun\,
    \peryear}}\,\frac{Q}{1.5\unit{MeV\,amu^{-1}}}\cgslum
  \label{eq:dch}
\end{equation}
\noindent where $Q$ is the average heat deposited in the NS crust per
accreted nucleon.

It was also suggested that the observed thermal spectrum of qLMXBs is
the result of the energy deposited in the crust, heating the NS core
during outbursts and thermally re-radiating away from the crust
through the NS atmosphere on core-cooling timescales \citep{brown98}.
It is thought that this atmosphere is composed of pure hydrogen since
heavy elements gravitationally settle on short timescale
($\sim$~seconds, \citealt{romani87,bildsten92}) once the accretion
from the low-mass evolved companion star onto the NS surface shuts off
after an outburst.  Models of NS H atmosphere \citep{rajagopal96,
  zavlin96, mcclintock04, heinke06}, now routinely used for qLMXBs,
imply emission-area radii consistent with the entire area of the NS,
compared to the $\approxlt 1\km$ emission-area radii in the previously
imposed blackbody approximation \citep{rutledge99}.  Spectral fitting
of qLMXBs with such models also leads to the determination of the
projected radius (as observed from infinity) defined by $\rinfty =
\rns g_{\rm r}^{-1} = \rns \left(1 - 2G\mns/\rns c^{2}\right)^{-1/2}$,
where \rns\ is the physical radius of the NS.

In the spectra of some qLMXBs in the field of the Galaxy (for example
Cen~X-4, \citealt{rutledge01a}, and Aql~X-1, \citealt{rutledge01b}),
an additional power-law component, dominating the spectrum above
$2\keV$ and unrelated to the H-atmosphere thermal emission, is
observed.  Proposed interpretations of this power-law tail include
residual accretion onto the NS magnetosphere
\citep{grindlay01a,cackett05}, shock emission via the emergence of a
magnetic field \citep{campana00b}, or an intrabinary shock between the
winds from the NS and its companion star \citep{campana04b}.  However,
recent analyses of the quiescent emission of LMXBs have shown that
variations in the non-thermal component are correlated to the
variations in the thermal component.  This suggests the presence of a
variable low-level accretion on the NS (for the LMXB XTE~J1701$-$462,
\citealt{fridriksson10}, and for the LMXB Cen~X-4, \citealt{cackett10}).

Another characteristic of qLMXBs is the expected lack of strong
variability on long or short timescales since their emission is
dominated by the thermal radiation from the interior of the NS.  While
the thermal component is not expected to display intensity
variability, other emission mechanisms (like those mentioned in the
previous paragraph) may be responsible for intensity and spectral
variations \citep{brown98}.  However, recent outburst episodes can
generate variations in the quiescent thermal luminosity on days to
years timescales, as observed for the LMXB KS~1731-260
(\citealt{rutledge02c}, following the models described in
\citealt{ushomirsky01}, see also \citealt{brown09}).

The DCH/H-atmosphere interpretation has been applied to a large number
of historical transient LMXBs and provided \rinfty\ measurements from
high signal-to-noise (S/N) spectra.  The following list is, to the
best of our knowledge, exhaustive: 4U~1608-522 \citep{rutledge99},
4U~2129+47 \citep{rutledge00}, Cen~X-4 \citep{campana00a,rutledge01a},
Aql~X-1 \citep{rutledge01b}, KS~1731-260 \citep{rutledge02c},
XTE~J2123$-$058 \citep{tomsick04}, EXO~1747$-$214 \citep{tomsick05},
MXB~1659$-$29 \citep{cackett06}, 1M~1716$-$315 \citep{jonker07a},
1H~1905+000 \citep{jonker07b}, 2S~1803$-$245 \citep{cornelisse07},
4U~1730$-$22 \citep{tomsick07}, EXO~0748$-$676 \citep{degenaar09}, and
XTE~J1701$-$462 \citep{fridriksson10}.

However, the 10\%--50\% systematic uncertainty in the distances to
qLMXBs in the field directly affects the \rinfty\ measurement
uncertainty.  Obtaining precise constraints on the dense matter
equation of state (EoS) is the observational motivation for measuring
the radii of NSs, requiring $\sim 5\%$ accuracy on \rinfty\ to be
useful for this purpose \citep[][]{lattimer04,steiner10}.  The known
distances to globular clusters (GCs) and their expected overabundances
of \xray\ binaries \citep{hut92} has motivated the search for qLMXBs
in the core of GCs.

There are 26 GC qLMXBs spectrally identified so far (see
\citealt{heinke03c} and \citealt{guillot09a} for two complementary
lists).  However, some of them have poorly constrained radius and
temperature measurements.  This can be due to low count statistics
and/or a large amount of galactic absorption (for example
$\nh=1.2\times10^{22}\unit{atoms\percmsq}$ for the GC Terzan~5) in
their direction which alters the low-energy end (0.1--1\keV) of the
spectra where the H-atmosphere spectrum of qLMXBs peaks.  Therefore,
their identification is regarded as less secure.  Most qLMXBs in GCs
have low-S/N spectra, and therefore, have rather large uncertainties
on their \rinfty\ measurements ($\sim15\%$ or more), precluding their
use for EoS constraints.  So far, only a few qLMXBs (in \OmCen,
\citealt{gendre03a}, in M13, \citealt{gendre03b}, and X7 in 47~Tuc,
\citealt{heinke06}) have spectra with S/N high enough to provide
useful constraints on the dense matter EoS \citep{webb07,steiner10}.

The close proximity of the globular cluster NGC~6397 ($d \approx
2.5\kpc$, \citealt{harris96, hansen07, strickler09}), the moderately
low galactic absorption\footnote {From
  http://cxc.harvard.edu/toolkit/colden.jsp using the NRAO data
  \citep{dickey90}.} ($\nh = 0.14\tee{22}\unit{atoms\percmsq}$, noted
$\nhtt=0.14$ hereafter) in its direction makes it a useful target for
the spectral identification of qLMXBs.  The discovery observation of
the qLMXB U24 \footnote{The source name U24 was used in the discovery
  paper (Gr01) and will be used throughout this work.} in NGC~6397
provided modest constraints on the NS projected radius: $\rinfty =
4.9\ud{14}{1}\km$ \citep[][Gr01 hereafter]{grindlay01b}.  The
proximity of U24 to the GC core requires the use of the \chandralong's
angular resolution to positionally and spectrally separate the qLMXB
from other sources in the crowded GC core.  U24 lies at $d_{\rm
  c}=6.8\,r_{\rm c} \approx 20 \arcsec$ from the GC center (core
radius $r_{\rm c} = 0.05\arcmin$ and half-mass radius $r_{\rm HM} =
2.33\arcmin$, NGC~6397 is a core-collapse cluster).  The reported
effective temperature of U24 was $\kteff = 57$--$92\eV$ (90\%
confidence interval, Gr01).

This paper presents the spectral and timing analyses of five archived
deep \chandra-ACIS observations of U24, located in GC NGC~6397.  These
long exposures provide the high-S/N data necessary to confirm the
qLMXB classification of U24 by obtaining precise \rns\ estimation.
The lack of variability also supports this classification.  The
organization of this paper is as follows.  Section~\ref{sec:red}
describes the data reduction and the analyses.
Section~\ref{sec:results} presents the results of the analyses.
Section~\ref{sec:discuss} provides a discussion of the results and
Section~\ref{sec:conclusion} is a short conclusion.

\begin{deluxetable}{rrrrr}[t]
  \tablecaption{\label{tab:Obs} \Chandra\ Observations of NGC~6397}
  \tablewidth{0pt}
  \tabletypesize{\scriptsize}    
  \tablecolumns{5}
  \tablehead{
    \colhead{Obs.} & \colhead{Starting} & \colhead{Exposure } &
    \colhead{Detector} & \colhead{Mode} \\
    \colhead{ID} & \colhead{Time (TT)} & \colhead{Time (ks)} &
    \colhead{} & \colhead{}  }
  \startdata 
  79   & 2000 Jul 31 15:31:33 &  48.34 & ACIS-I3 (FI) &  F \\
  2668 & 2002  May 13 19:17:40 &  28.10 & ACIS-S3 (BI) &  F \\
  2669 & 2002  May 15 18:53:27 &  26.66 & ACIS-S3 (BI) &  F \\
  7460 & 2007 Jul 16 06:21:36 & 149.61 & ACIS-S3 (BI) & VF \\
  7461 & 2007 Jun 22 21:44:15 &  87.87 & ACIS-S3 (BI) & VF \\
  \enddata
  \tablecomments{All observations were performed with a 3.24104\sec\
  frame rate.  TT refers to terrestrial time.  FI and BI refer to
  front-illuminated and back-illuminated detectors, respectively.}
\end{deluxetable}

\section{Data Reduction and Analysis}
\label{sec:red}

\begin{deluxetable*}{ccccccc}[t]
  \tablecaption{\label{tab:PosU24} Position of U24 in \chandra\ 
    Observations of NGC~6397 }
  \tablewidth{0pt}
  \tabletypesize{\scriptsize}    
  \tablecolumns{6}
  \tablehead{
    \colhead{Obs. ID} & \colhead{R.A.} & \colhead{$\Delta_{\rm R.A.}$} & 
    \colhead{Decl.} & \colhead{$\Delta_{\rm Decl.}$} & \colhead{Detection} &
    \colhead{Reference} \\
    \colhead{} & \colhead{(J2000)} & \colhead{(\arcsec)} & \colhead{(J2000)} &
    \colhead{(\arcsec)} & \colhead{Significance} & \colhead{}  }
  \startdata 
      79 & 17$^{\rm h}$40$^{\rm m}$41.421$^{\rm s}$ & 0.02 & --53\deg 40\arcmin 04.73 & 0.02 &     --      & Gr01 \\
      79 & 17$^{\rm h}$40$^{\rm m}$41.459$^{\rm s}$ & 0.6  & --53\deg 40\arcmin 04.47 & 0.6  & 128$\sigma$ & This work\\
    2668 & 17$^{\rm h}$40$^{\rm m}$41.489$^{\rm s}$ & 0.6  & --53\deg 40\arcmin 04.38 & 0.6  & 152$\sigma$ & This work\\
    2669 & 17$^{\rm h}$40$^{\rm m}$41.485$^{\rm s}$ & 0.6  & --53\deg 40\arcmin 04.53 & 0.6  & 149$\sigma$ & This work\\
    7460 & 17$^{\rm h}$40$^{\rm m}$41.486$^{\rm s}$ & 0.6  & --53\deg 40\arcmin 04.60 & 0.6  & 302$\sigma$ & This work\\
    7461 & 17$^{\rm h}$40$^{\rm m}$41.488$^{\rm s}$ & 0.6  & --53\deg 40\arcmin 04.54 & 0.6  & 228$\sigma$ & This work\\
  \enddata 
  
  \tablecomments{$\Delta_{\rm R.A.}$ and $\Delta_{\rm Decl.}$ are the
    uncertainties on the position, dominated by the
    \Chandra\ systematic uncertainty in this work.  The positions
    reported previously \citep[][, noted Gr01]{grindlay01b} have been
    corrected for the \Chandra\ systematic uncertainty (see details in
    Section~\ref{sec:pos}).}
\end{deluxetable*}

\subsection{Observations, Source Detection, and Count Extraction}
\label{sec:obs}

We analyze one archived \chandra/ACIS-I and four archived
\chandra/ACIS-S observations of NGC~6397 (Table~\ref{tab:Obs}).  The
source detection and the data analysis are performed using the CIAO
v4.1.1 package \citep{fruscione06}.  The event files (level=1) are
re-processed with the latest calibration files from CALDB v4.1
\citep[with the latest effective area maps, quantum efficiency maps,
  and gain maps]{graessle07}, as recommended in the CIAO Analysis
Thread ``\emph{Reprocessing Data to Create a New Level=2 Event File}''
to include the up-to-date CTI corrections (charge-transfer
inefficiency).

The re-processed event files are analyzed including counts in the
0.5--8.0\keV\ energy range.  The full-chip light curves do not show
evidence of background flares in any of the five observations,
allowing use of the full exposure time of each observation.  The low
flux of the source of interest, U24, allows neglecting pile-up; the
count rate of $\sim 0.06$ photons per frame corresponds to a pile-up
fraction of less than 2\% \footnote{\Chandra\ Observatory Proposer
  Guide v12.0, Figure~6.18, December 2009}.

For each observation (ObsID), the source detection is performed with
the CIAO {\tt wavdetect} algorithm.  An exposure map is created using
the task {\tt mkexpmap} prior to the source detection.  The {\tt
  wavdetect} exposure threshold \emph{expthresh} is set to 0.1 and the
wavelet detection scales are set to \emph{scales = ``1.0 2.0 4.0
  8.0''}.  Thirty-five sources are detected ($\sigma>3$) on the ObsID
2668, 37 sources on ObsID 2669, 66 sources on ObsID 7460, 48 on ObsID
7461, and 37 on the ACIS-I observation ObsID 79.

This paper is solely focused on radius measurement and timing analysis
on the qLMXB U24.  While the source detection is performed over the
whole ACIS chip, the following analysis pertains only to the source
U24.

Counts are extracted with the CIAO script {\tt psextract} around the
source position in a circular region of radius 3\arcsec, which ensures
that 98\% of the enclosed energy fraction at 1\keV\ is included
\footnote{\Chandra\ Observatory Proposer Guide v12.0, Figure~6.7,
  December 2009}.  The closest detected source, located at
10.6\arcsec\ distance from U24, has 21.6 counts (background
subtracted) within 1.5\arcsec.  It contributes to $\ll0.04$
contamination counts within the extraction radius of U24 (on the
longest observation).  The background is extracted from an annulus
centered at the qLMXB position with inner radius 5\arcsec, and outer
radius 30\arcsec.  Other detected sources within the background
annulus are excluded with a 5\arcsec\ radius region, which eliminate
99.8\% of source counts in the background region.  For the deepest
observation (ObsID 7460), 15 counts from other sources are within the
extracted background (which contains 6187 counts).  In other words,
these constraints ensure that $\sim 0.25\%$ of the background counts
are due to other sources.  Finally, the extraction radius (containing
98\% of the ECF) does not require to apply a correction to the flux.

Following the CIAO Science Thread \emph{``Creating ACIS RMFs with
  mkacisrmf''}, the response matrices files (RMFs) are recalculated
prior to the spectral analysis since the RMFs obtained from {\tt
  psextract} are not suited for ACIS observations with focal plane
temperature of --120\deg C (the usual {\tt mkrmf} command does not use
the latest calibration available in the case of --120\deg C ACIS
imaging data).  In addition, the ancillary response files (ARFs) are
also recalculated using the energy grid of the newly obtained RMFs.
Overall, the extracted spectra, together with the RMFs and ARFs, are
used for the spectral analysis.  In those spectra, the effect of
background counts can be ignored.  Indeed, in the worst case (for
ObsID 7460), the number of expected background events accounts for
2.4\% of the total number of counts in the extracted region (78.0
background counts out of a total of 3188 counts), so that the
background is neglected for the spectral analysis.

\subsection{Spectral Analysis}
\label{sec:spectra}

For each of the five observations, two spectral files are created, one
with unbinned events (for fitting with the Cash-statistics,
\citealt{cash79}) and one with binning (for fitting with the
\chisq-statistics).  For the latter, the bin width in the
0.5--1.5\keV\ energy range matches the energy resolution of the
ACIS-S3 chip, i.e., $\sim$0.15\keV.  Above 1.5\keV, four wider bins
(0.3\keV, 0.6\keV, and two 3\keV\ wide spectral bins) are created.  In
some cases of low count statistics, the last 2 or 3 bins are grouped
together to maintain a minimum of 20 counts per bin.  The main
criterion for the creation of the spectral bins is the energy
resolution of the detector, but the 20 counts minimum is imposed to
ensure approximate Gaussian uncertainty in each bin.  Such a binning
avoids an artificially small reduced-\chisq, and conserves the
validity of \chisq-statistics.

Spectral fitting is performed with the software \emph{XSPEC} v12.5.1 
\citep{arnaud96} using the publicly available model of NS H-atmosphere
{\tt nsatmos} \citep{mcclintock04, heinke06}.  The model assumes
non-magnetic NSs and has been computed for a range of surface gravity
$g=(0.1-10)\tee{14} \cgsaccel$.  For the normalization parameter, {\tt
  nsatmos} uses the emitting fraction of the NS surface.  It is kept
fixed to unity in this work; in other words, the whole NS surface
emits.  The distance parameter is held fixed as well at the value of
NGC~6397, $d=2.5\kpc$ \citep{hansen07,strickler09}.  The NS mass is
assumed to be $1.4\msun$.  Finally, the galactic absorption is taken
into account using the {\tt phabs} multiplicative model, with \nh, the
hydrogen column density parameter, set to $\nhtt = 0.14$.  The errors
on the best-fit parameters (\rns, \kteff) are calculated using the
command {\tt error} in \emph{XSPEC} with 90\% confidence or using the
command {\tt steppar}.  Confidence contours in mass--radius space are
obtained with the {\tt steppar} command with both the mass and the
radius as free parameters.  The results of the spectral analysis are
presented in Section~\ref{sec:spec_res}.

\subsection{Variability Analysis}
\label{sec:variability}

For each of the five observations, we perform two analyses to search
for source variability on timescales shorter than the duration of the
observations, and one analysis for timescales spanning the time
between the first and last observations.

{\bf Power Density Spectrum (PDS)}.  The data from each observation
were binned into a discrete light curve, with time bin size equal to
the time resolution used in the observation ($\Delta T=3.24104\sec$).  A
standard fast-Fourier transform (FFT) of the discrete data into
frequency space was produced \citep{press95}, using an open-source FFT
algorithm \citep{frigo98}. This produced a Fourier transform of the
data, covering the frequency range $1/T-0.15427\unit{Hz}$, where $T$
is the duration of the observation, and $0.15427\unit{Hz}$ is the
Nyquist frequency, with discrete frequency resolution $1/T$, and a
total of $N/2$ frequency bins, where $N=T/\Delta T$ is the number of
time bins in the light curve.  The resulting Fourier transform was
then converted into a PDS, where the power $P_j$ in each frequency bin
$j$ is $P_j= a_j^2 + b_j^2$, where $a_j$ and $b_j$ are the real and
imaginary parts of the Fourier component associated with a frequency
$f_j=j/T$, producing a PDS.  The data were then normalized according
to a standard prescription for analyses \citep{leahy83}.

{\bf Short-Timescale ($<$1\,day) Variability}.  Short-term variability
in each ObsID is assessed by visual inspection of the five sources
light curves.  In addition, a Kolmogorov--Smirnov (K-S) test
\citep{press95} is performed to quantify the consistency of the
temporal distribution of counts in each observation with a constant
count rate.  More specifically, the integrated-count light curve is
compared to a linear distribution using a K-S test.  A low K-S
probability would indicate that the integrated light curve is
significantly different from a linear distribution, demonstrating the
presence of variability on the timescale of the observation.

{\bf Long-Timescale ($\sim$months-years) flux variability}.  Long-term
variability is investigated to determine if the flux remained constant
over the course of the five observations, between 2000 and 2007.  This
is performed by adding a multiplicative constant parameter to the
spectral model, and by fitting the five spectra simultaneously.  The
constant is kept fixed at a value $c=1$ for one spectrum (ObsID 79)
and as a free and untied parameter for the remaining four spectra
while the remaining spectral parameters (\nh, \rns, \kteff) are
assumed to be the same across all observations.  Best-fit $c$ values
statistically consistent with unity would demonstrate that the source
flux remained constant on the timescale of the five observations.

\section{Results}
\label{sec:results}

\begin{figure}[t]
  \centerline{~\psfig{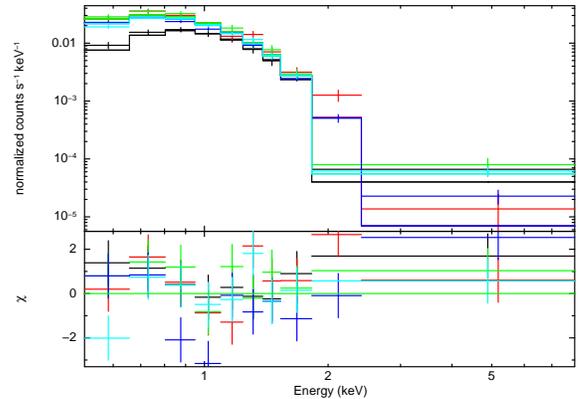}~}
  \bigskip
  \caption[]{Folded spectra of all five \chandra-ACIS observations of
    U24 in NGC~6397, in the 0.5--8.0\keV\ energy range.  Each color
    corresponds to the spectra of each observations: ObsID 79 in
    black, ObsID 2668 in red, ObsID 2669 in green, ObsID 7460 in blue,
    ObsID 7461 in cyan blue [See the electronic edition of the paper
      for a color version of this figure].  The solid lines are the
    best-fitting model of an NS H-atmosphere model: {\tt NSATMOS},
    with $\nhtt=0.14$ and $M_{\rm NS} = 1.4\msun$ kept frozen.  The
    spectral binning was performed in accordance with the
    \chandra\ energy resolution (see Section~\ref{sec:spectra}).  The
    fit is statistically acceptable with \Chisq{1.49}{45}{0.02}.  The
    residuals, shown in the lower part, indicate evidence of a count
    deficiency between 0.8\keV\ and 1.2\keV\ (see
    Figure~\ref{fig:spec7460} and Section~\ref{sec:spec_res}), as well
    as an apparent excess of counts at large energy, above 2.5\keV,
    which is fit with a power-law component with $\alpha = 1$ to
    estimate the contribution to the total flux (see
    Section~\ref{sec:spec_res}). \label{fig:allspec}}
\end{figure}

\subsection{Positional Analysis}
\label{sec:pos}

The position of the source reported in the discovery observation
(ObsID 79) is R.A.=17$^h$40$^m$41.421$^s$ and ${\rm decl.}=-53\deg
40\arcmin 04.73\arcsec$ (Gr01).  The authors corrected for the
systematic \chandra\ uncertainty in the pointing (0.6\arcsec) by
cross-identifying cataclysmic variables on \chandra\ and
\hstlong\ (HST) observations.  The positional uncertainty on the
source is subarcseconds and consists in the residual error from the
correction and the statistical uncertainty from the detection
algorithm.

For comparison, the source detection in all five observations
presented here led to the positions listed in Table~\ref{tab:PosU24}.
It is concluded that the position of U24 is consistent with that
reported previously, within the 0.6\arcsec\ \chandra\ systematic
uncertainty.  No correction for the systematic uncertainty is
performed here since the identification of U24 is free of source
confusion and since the purpose of this paper focuses on the spectral
analysis.

\subsection{Spectral Analysis}
\label{sec:spec_res}

\begin{deluxetable*}{ccccccc}[h]
  \tablecaption{\label{tab:results} Spectral Parameters of U24 in \chandra\ 
    Observations of NGC~6397 }
  \tablewidth{0pt}
  \tabletypesize{\scriptsize}    
  \tablecolumns{6}
  \tablehead{
    \colhead{Obs. ID} & \colhead{\nhtt} & \colhead{\rns\ (\km)} & \colhead{\kteff\ (\eV)} & 
    \colhead{$F_{\rm X}$} & \colhead{Statistics}  & \colhead{PL Contrib.}}
  \startdata 
       79 & (0.14)              &10.7\ud{3.3}{2.3}  & 73\ud{10}{10} & 1.50\ud{0.14}{0.22} & Cash: 256.7 (100\%) & -- \\
          & 0.09\ud{0.03}{0.03} & 6.8\ud{5.1}{1.8}  & 97\ud{38}{32} & 1.27                & Cash: 254.2 (0\%)   & -- \\
          & (0.14)              &10.1\ud{3.4}{5.1p} & 75\ud{49}{11} & 1.49\ud{0.11}{0.22} & \Chisq{0.76}{7}{0.62} &  $\leq$ 8.8\%\\
          & 0.10\ud{0.05}{0.03} & 7.2\ud{3.2}{2.2p} & 93\ud{42}{28} & 1.30                & \Chisq{0.60}{6}{0.73} &  --\\
     \hline
     2668 & (0.14)              & 7.2\ud{2.4}{1.8}  & 95\ud{28}{19} & 1.47\ud{20}{1.34}   & Cash: 213.3 (100\%) & -- \\
          & 0.13\ud{0.03}{0.02} & 7.1\ud{2.8}{1.6}  & 96\ud{26}{21} & 1.44                & Cash: 213.1 (0\%)   & -- \\
          & (0.14)              & 7.8\ud{3.4}{2.8}  & 88\ud{43}{19} & 1.44\ud{1.20}{5.81} & \Chisq{1.81}{9}{0.61} & $\leq$ 3.9\%\\
          & 0.13\ud{0.04}{0.03} & 7.2\ud{4.6}{2.2}  & 94\ud{41}{27} & 1.38                & \Chisq{1.97}{8}{0.46} & -- \\
     \hline
     2669 & (0.14)              &10.0\ud{2.6}{5.0p} & 76\ud{37}{9}  & 1.52\ud{0.06}{0.26} & Cash: 159.0 (98\%) & -- \\
          & 0.11\ud{0.05}{0.02} & 7.0\ud{5.4}{2.0p} & 96\ud{41}{32} & 1.36                & Cash: 157.7 (0\%) & -- \\
          & (0.14)              & 9.9\ud{3.0}{4.9p} & 77\ud{51}{10} & 1.51\ud{0.05}{0.40} & \Chisq{0.63}{8}{0.74} & $\leq$ 4.2\%\\
          & 0.12\ud{0.07}{0.03} & 7.2\ud{8.8}{2.2p} & 93\ud{44}{31} & 1.37                & \Chisq{0.66}{7}{0.70} & --\\
     \hline
     7460 & (0.14)              & 9.9\ud{1.2}{1.2}  & 75\ud{6}{5}   & 1.35\ud{0.04}{0.09} & Cash: 367.0 (100\%) & -- \\
          & 0.11\ud{0.01}{0.01} & 6.9\ud{1.4}{0.9}  & 94\ud{13}{12} & 1.18                & Cash: 355.1 (0\%) & -- \\
          & (0.14)              &10.1\ud{1.5}{1.5}  & 74\ud{7}{6}   & 1.35\ud{0.54}{0.17} & \Chisq{2.32}{8}{0.02} & $\leq$ 4.7\%\\  
          & 0.11\ud{0.02}{0.01} & 7.0\ud{1.9}{1.5}  & 93\ud{25}{21} & 1.19                & \Chisq{2.14}{7}{0.04} & --\\ 
     \hline
     7461 & (0.14)              & 6.7\ud{1.9}{1.0}  &100\ud{19}{17} & 1.37\ud{5.29}{1.19} & Cash: 291.0 (0\%) & --\\
          & 0.14\ud{0.02}{0.02} & 6.6\ud{2.6}{1.1}  &101\ud{20}{24} & 1.38                & Cash: 291.0 (0\%) & -- \\
          & (0.14)              & 6.1\ud{3.4}{0.9}  &104\ud{24}{31} & 1.37\ud{1.33}{1.07} & \Chisq{0.97}{7}{0.45} & $\leq$ 3.3\%\\
          & 0.17\ud{0.06}{0.04} &10.7\ud{5.3}{5.7p} & 74\ud{52}{14} & 1.57                & \Chisq{1.03}{6}{0.41} & --\\
     \hline
    All 5 & (0.14)              & 8.9\ud{0.9}{0.6}  & 80\ud{4}{5}   & 1.39\ud{0.02}{0.06} & Cash: 1289.9 (100\%)  & -- \\
          & 0.12\ud{0.01}{0.01} & 6.9\ud{1.0}{0.7}  & 95\ud{9}{10}  & 1.28                & Cash: 1276.1 (69.2\%) & -- \\
          & (0.14)              & 9.3\ud{1.0}{1.0}  & 78\ud{5}{4}   & 1.39\ud{0.04}{0.06} & \Chisq{1.49}{45}{0.02} & $\leq$ 3.7\%\\
          & 0.12\ud{0.02}{0.01} & 7.2\ud{1.9}{1.5}  & 92\ud{22}{14} & 1.29                & \Chisq{1.16}{44}{0.03}& --\\
  \enddata 

  \tablecomments{\tiny{The model used for all spectral fits presented
    here is the NS H-atmosphere model {\tt NSATMOS}, with the distance
    to the source fixed at the value $d=2.5\kpc$ and the mass of the
    NS fixed at $M_{\rm NS}=1.4\msun$.  \nhtt\ is the galactic
    absorption \nh\ in units of $\ee{22}\unit{atoms\percmsq}$. Values
    in parentheses are held fixed for the spectral fits and quoted
    uncertainty are 90\% confidence .  $F_{\rm X}$ is the unabsorbed
    flux in units of $\ee{-13}\cgsflux$ in the 0.5--10\keV\ range.  The
    uncertainties on the fluxes were only calculated for the
    fixed-\nh\ fits.  For ObsID 7460, the \chisq\ value greater than
    2 prevents \emph{XSPEC} to directly estimate the error region.  It
    was obtained using the command {\tt steppar}.  ``PL contrib.''
    refers to the upper limit of a power-law component contribution to
    the total flux (see Section~\ref{sec:spec_res}). The percent in
    parenthesis following the Cash-statistic value indicates the
    goodness-of-fit.  The best fit parameters and the statistics
    information provided do not include the power-law component.  }}
\end{deluxetable*}

The spectral analysis is performed using the Cash-statistics on the
unbinned data (neglecting the background) and using \chisq-statistics
on the grouped data.  Results are presented in
Table~\ref{tab:results}, listing the best-fit parameters for Cash and
\chisq-statistics, freezing and thawing \nh.  All reported errors are
90\% confidence.  The \chisq-statistics null hypothesis probability
confirms the viability of the fitted model while the Cash-statistics
assumes that the model describes the data.  Nonetheless, the use of
Cash-statistics provides smaller uncertainties on the best-fit
parameters, specifically for the NS radius measurements.

A simultaneous fit is also performed using all five spectra, therefore
increasing the count statistics and providing better constraints on
the best-fit parameters (Figure~\ref{fig:allspec}).  Prior to that,
the source long-term variability is inspected, as described in
Section~\ref{sec:variability}.  The best-fit values (with
Cash-statistic) for the multiplicative constants described in
Section~\ref{sec:variability} are $c=1.02\ud{0.09}{0.10}$,
$c=1.04\ud{0.09}{0.10}$, $c=0.92\ud{0.06}{0.07}$, and
$c=0.95\ud{0.07}{0.08}$, for the observations ObsID 2668, ObsID 2669,
ObsID 7460, and ObsID 7461, respectively, and $c=1$ (fixed) for ObsID
79.  All best-fit values are statistically consistent with unity
(within $1.5\sigma$), indicating that the source flux did not vary on
long-term timescale.  This allows for simultaneous spectral fitting
with the constant multiplier fixed at the value $c=1$ for all five
spectra.

The upper limit of a power-law contribution to the total flux is also
estimated.  To do so, a power-law component with fixed photon index
$\alpha = 1$ is added to the NS atmosphere model and the flux of this
component using the upper limit of the power-law normalization
parameter is measured.  It is found that the power-law contribution
accounts for $\leq 3.7\%$ of the total flux (90\% confidence upper
limit), when estimated from the simultaneous fits.  Upper limits on
the contribution of a power-law component for the individual spectra
are also indicated in Table~\ref{tab:results}.

\begin{figure}[h]
  \centerline{~\psfig{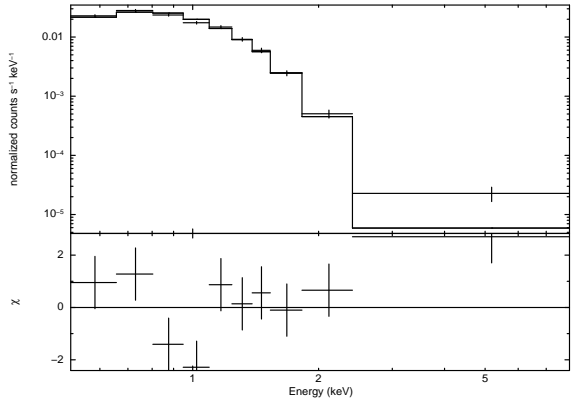}~}
  \bigskip
  \caption[]{Folded spectrum of the ACIS observation 7460.  As for
    Figure~\ref{fig:allspec}, the solid line shows the best-fitting
    {\tt NSATMOS} model.  The count deficiency observed between
    0.8\keV\ and 1.2\keV\ does not exceed the 99\% limit from the MC
    simulations of 10,000 spectra.  In other words, the detection is
    not significant enough to claim the presence of an absorption line
    (see Section~\ref{sec:spec_res}). \label{fig:spec7460}}
\end{figure}

There is some evidence for a count deficiency between 0.8 and
1.2\keV\ on ObsID 7460 given the assumed spectrum and instrument
calibration (see Figure~\ref{fig:spec7460}).  We parameterize this
apparent dip in the spectrum with a {\tt notch} component.  The
best-fit {\tt notch} central energy is $E=0.96\ud{0.03}{0.02}\keV$,
and the best-fit width is $W=42\pm17\eV$.  The added component
improves the statistics, \Chisq{1.35}{6}{0.23}, without altering the
best fit NS temperature and radius.

\begin{figure}[t]
  \centerline{~\psfig{file=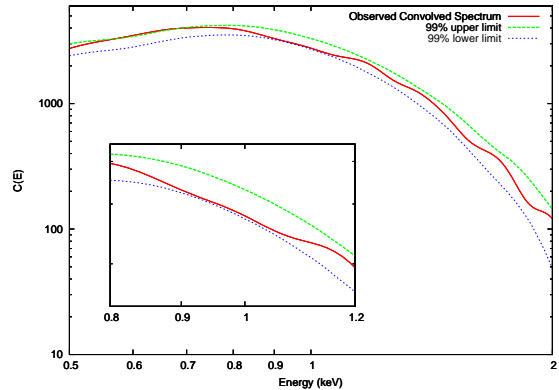,width=7.5cm,angle=-90}~}
  \bigskip
  \caption[]{Results of the MC simulations analyzing the count
    deficiency observed in Figure~\ref{fig:spec7460}, in the energy
    range 0.8--1.2\keV.  The observed spectrum (solid line) convolved
    with the ACIS energy redistribution, $C\left(E\right)$, is shown
    within the 99\% upper- and lower-limit envelopes (dashed and
    dotted lines, respectively) obtained from Monte Carlo simulations
    (see Section~\ref{sec:spec_res}).  \label{fig:envelope}}
\end{figure}

We also investigate the statistical significance of this deficiency.
Using the method described in a previous work \citep{rutledge03}, we
estimate the probability of observing such a deviation from a
continuum model.  First, the observed spectrum is convolved with the
energy redistribution of the ACIS-S detector.  Then, using a
Monte Carlo approach, it is shown that the convolved spectrum does not
exceed the 99\% confidence limits envelope obtained from 10,000
simulated spectra (Figure~\ref{fig:envelope}).  They were created
using the best-fit {\tt nsatmos} model ($\rns = 8.9\km$, $\kteff =
80\eV$, $\mns = 1.4$, $d=2.5\kpc$, and $\nhtt=0.14$, see
Table~\ref{tab:results}).  The maximum deviation corresponds to a
98.2\% confidence, which does not constitute sufficient evidence to
claim the detection of an absorption line, but which should be
investigated in more details with higher S/N observations.

For completeness, the results of spectral fits with other models are
provided.  Using an absorbed simple blackbody model, the fit to the
spectra (all five binned spectra) is not statistically acceptable:
\Chisq{3.51}{45}{$2\tee{-14}$} with \nhtt=0.14 fixed, and
\Chisq{1.631}{44}{$5\tee{-3}$} with \nhtt\ allowed to vary.  A thermal
bremsstrahlung model is also fit to the spectra, leading to a
statistically acceptable fit (\Chisq{1.45}{45}{0.03}) with best-fit
parameter $kT = 391\pm10\eV$.

\subsection{Variability Analyses}

We find no evidence of broadband variability as a function of
frequency in the PDS.  For broadband variability, uncertainties were
assigned to each frequency bin equal to the square root of the
theoretical variance in the power, appropriate to the assumed
normalization \citep{leahy83}, and derived 90\% confidence ($2\sigma$)
upper-limits on the root-mean-square (rms) variability.  The data were
then rebinned logarithmically and the resulting PDS were fit with a
model of a constant power \footnote{The constant accounts for the
  expected Poisson noise power in the PDS, which is approximately 2 in
  the PDS normalization used; however, we observed the Poisson level
  for each of the PDS to be suppressed, to a value of $\sim1.9$, due
  to instrumental dead-time effects.  To correct for this, on average,
  in the rms variability upper limits derived here, the measured rms
  limits were increased by a factor $\langle A \rangle/2.0$, where
  $\langle A \rangle$ is the best-fit Poisson power level.}, plus a
power-law component in which the power scales $P_j\propto
f_j^{-\alpha}$, and the power-law slope was held fixed at $\alpha=1$.
This model was fit to the data using a Levenberg--Marquardt
\chisq\ minimization technique \citep{press95}, to find the best-fit
model parameters.  The resulting 90\% confidence upper limits on the
broadband variability, in a frequency range $0.0001$--$0.10\unit{Hz}$
(used for each observation, to ease comparison of limits, although the
longer observations are sensitive to variability at frequencies below
this range) and across the full \chandra/ACIS energy range
(0.5--8.0\keV) were : $<12\%$ (ObsID 79), \chisqrnu=0.63 (26 dof);
$<19\%$ (ObsID 2668), \chisqrnu=1.5 (27 dof);$<11\%$ (ObsID 2669),
\chisqrnu=1.32 (26 dof); $<$6.4\% (ObsID 7460), \chisqrnu=0.98 (15
dof); $<$8.6\% (ObsID 7461), \chisqrnu=0.98 (9 dof).

We find no evidence of periodic variability.  For ObsIDs 79, 2668,
2669, 7460, and 7461 respectively, we find maximum (normalized) powers
of $P_{\rm max}=$21.8, 17.0, 15.3, 17.7, and 23.7 which, with a number
of frequency bins of 7526, 4383, 4154, 23072, and 13812, correspond to
respective probabilities of chance occurrence in all cases of $>$10\%.

Visual inspection of the light curves did not reveal any variation in
the source count rates.  Short-term variability was also investigated
in a more quantitative way by comparing the integrated ligthcurve with
a linear distribution.  None of the five observations showed an
integrated light curve significantly different from a linear
distribution.  More specifically, the calculated K-S probabilities
were 65\% (ObsID 79), 62\% (ObsID 2668), 95\% (ObsID 2669), 51\%
(ObsID 7460) and 16\% (ObsID 7461).  Therefore, we find no evidence of
intensity variability on the timescale of the integration times,
i.e., $\approxlt 1\unit{day}$.

\section{Discussion}
\label{sec:discuss}

\subsection{\rinfty\ Calculation}

\begin{figure}[t]
  \centerline{~\psfig{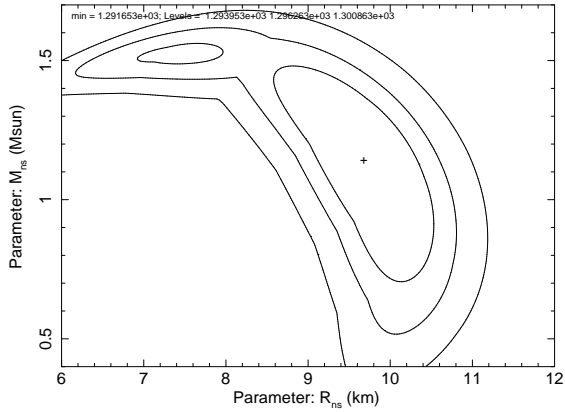}~}
  \bigskip
  \caption[]{Contour plot in mass--radius space resulting from the
    simultaneous fit of the five data sets with the {\tt NSATMOS}
    model.  The 67\%, 90\%, and 99\% contours are shown, and the cross
    indicates the best fit (when both the mass and the radius are free
    to vary).  For an NS mass of 1.4\msun, the range of \rns
    corresponds to that obtained and quoted in
    Section~\ref{sec:spec_res}. \label{fig:contours}}
\end{figure}

Producing realistic constraints on the dEoS requires obtaining values
of \rinfty\ relying on as few assumptions as possible.  Keeping the
mass fixed for the model fitting is therefore not appropriate for that
purpose.  We estimate the value of \rinfty\ by permitting \mns\ to
vary, and calculating the contours of constant model probability
resulting from the fits in a mass--radius space
(Figure~\ref{fig:contours}).  This is done using the {\tt steppar}
command in \emph{XSPEC}.  The 90\% confidence regions of \rns\ and
\mns\ are obtained from the 90\%-contour in the mass--radius space:
$\rns = 9.7\ud{0.9}{0.8}\km$ and $\mns = 1.13\ud{0.47}{0.32}\msun$.
The best-fit value of the projected radius is therefore
$\rinfty=11.9\km$.  The calculation of the uncertainties is
complicated by the fact that the distribution of \rns\ and \mns\ is
not symmetric around the best-fit values (i.e., not Gaussian).
Moreover, \rns\ and \mns\ are highly correlated, as shown by the
crescent shape of the contour in M-R space
(Figure~\ref{fig:contours}).  Therefore, the calculation of the
uncertainties on \rinfty\ using Gaussian normal error propagation is
not valid.  We describe two methods to obtain the uncertainty on
\rinfty.

The projected radius and its uncertainties can be obtained from a
tabulated version of the {\tt nsa} spectral model \citep{zavlin96}.
However, this model (or {\tt nsa}) is less adapted than the {\tt
  nsatmos} or {\tt nsagrav} models \citep{webb07} because it was
calculated for a single value of the surface gravity $g =
2.43\tee{14}\cgsaccel$ while the other two models consider a range of
values.  Nevertheless, the best-fit \rinfty\ value with this model is:
$\rinfty= 12.1 \ud{1.5}{0.9}\km$ (consistent with the value calculated
in the previous paragraph), for a temperature $\kteff=76\ud{2}{3}\eV$
(\Chisq{1.54}{44}{0.02}).

A second method to estimate the uncertainties involves geometric
construction, by reading graphically the error region of \rinfty\ on
the M-R contours (Figure~\ref{fig:contours}).  For that, we choose to
use the line of constant surface gravity (i.e., constant
$M\left(R\right)$) that goes through the point $\left(R,M\right) =
\left(0\km,0\msun\right)$ and the point of best fit in M-R space.
This line intersects the 90\% contour at the points $\left(R,M\right)
= \left(9.048\km,1.045\msun\right)$ and $\left(R,M\right) =
\left(10.39\km,1.245\msun\right)$.  These two points correspond to the
values $\rinfty=11.15\km$ and $\rinfty=12.92\km$ which are,
respectively, estimates of the lower and upper 90\% confidence
uncertainties on $\rinfty$, assuming a constant value of the surface
gravity.

Therefore, the projected radius and its estimated 90\% confidence
uncertainties are: $\rinfty=11.9\ud{1.0}{0.8}\km$.  With the achieved
uncertainty, U24 becomes the third best radius measurement of a NS
among the population of GC qLMXBs, after the ones in
\OmCen\ \citep{gendre03a} and in M13 \citep{gendre03b}.

\subsection{Error budget}
The high S/N spectra and the precise radius measurements obtained in
the work presented here can be used to constrain the EoS of dense
matter.  A high precision on the NS radius is mandatory to exclude
some of the existing nuclear dense matter EoSs and provide the
necessary constraints to understand the behavior of such matter.
However, other sources of error come into play in this type of
measurements.  To quantify the total uncertainty on the radius
measurement presented here, we estimate the contribution of each
source of error into an error budget, including the distance to the GC
NGC~6397, uncertainties intrinsic to the model used, systematic and
statistical uncertainties. In those references where these
uncertainties are discussed (for example, \citealt{heinke06}), only
two of the three uncertainties we discuss here (distance and detector
systematics) are addressed.  No work that we can find in the
literature discusses the impact of the uncertainty in the spectral
model on derived model parameters; therefore, we do so here.

\begin{itemize}
\item The distance to the GC was recently measured using two
  independent methods.  The analysis of the CO white dwarf (WD)
  sequence from deep observations in an outer field of NGC~6397 led to
  a distance of 2.54\ppm0.07\kpc\ \citep{hansen07}.  More recently,
  using CO WDs in central regions of the cluster, the distance was
  calculated to be 2.34\ppm0.13\kpc\ \citep{strickler09}.  The
  weighted mean of these two measurements is 2.50\ppm0.06\kpc,
  corresponding to a 2.4\% uncertainty.  The unknown line-of-sight
  position of U24 within NGC~6397 accounts for $<0.1\%$ of the
  distance uncertainty, which can be neglected compared to the GC
  distance uncertainty.

\item The calculation of spectral model {\tt NSATMOS} also contributes
  to the total uncertainty on the measured radius.  However, the
  previous works describing the model do not provide a discussion on
  the fractional uncertainties in intensity due to convergence during
  the calculation of the spectral model \citep{mcclintock04,heinke06}.
  Therefore, it is not possible to evaluate the errors of the
  resulting spectra. The cited reference for similar models
  \citep[{\tt NSA} and {\tt NSAGRAV},][]{zavlin96} only provides
  information on the temperature calculation convergence, which does
  not permit an estimation of the uncertainty error in the modeled
  intensity as a function of energy.

\item The statistical uncertainties are those quoted in
  Table~\ref{tab:results} (90\% confidence).  This includes the 3\%
  systematic uncertainty of the detector calibration, taken into
  account using the ``{\tt systematic 0.03}'' command in \emph{XSPEC}.

\end{itemize}

Consequently, the distance uncertainty (2.4\%) is the only
quantifiable error not taken into account in the radius measurement
obtained from spectral fitting.  It is therefore added in quadrature
to the systematic and statistical uncertainties to obtain the total
quantifiable uncertainty in the radius measurement.  For example, the
upper bound uncertainty limit of \rns\ was 10.1\% and is 10.4\% when
accounting for the distance uncertainty.  The lower bound uncertainty
limit was 6.7\% and becomes 7.1\%.  Consequently, the physical radius
is $\rns = 8.9\ud{0.9}{0.6}$ (for $\mns = 1.4 \msun$) while the
estimated radiation radius is $\rinfty = 11.9\ud{1.0}{0.8}\km$, when
considering the sources of uncertainty listed above.  In conclusion,
in NGC~6397, the distance uncertainty of NGC~6397 alone minimally
affects the current uncertainty on the radius measurement.

\subsection{Core temperature calculation}
The best-fit temperature and physical radius of the NS can be used to
determine the interior temperature.  This calculation is
model dependent and due to the uncertainties in the deep atmosphere
composition (the depth of the H/He transition in particular), two
different models are considered here.  The first one assumes a layer
of helium down to a column depth $y=1\tee{9}\cgsdepth$ with a pure
layer of iron underneath.  The second model considers a thin layer of
He down to $y=1\tee{4}\cgsdepth$ with a mixture dominated by
rp-processes.  These two alternatives take into account the extremal
values for the core temperature, for a fixed effective temperature.

The calculation, described in a previous work \citep{brown02}, was
performed with a fixed mass $\mns = 1.4 \msun$ and a fixed radius
$\rns = 8.9\km$, which corresponds to the best-fit value.  The
effective temperature used was $\kteff = 80 \ud{4}{5}\eV$.  The
calculation is performed down to a column depth of
$1\tee{14}\cgsdepth$, since the temperature is nearly isothermal in
deeper layers.

For the first model, the resulting interior temperature (at
$y=\ee{14}\cgsdepth$) is $T_{\rm core} = \left( 3.37\ud{0.36}{0.41}
\right) \tee{7}\unit{K}$.  The second model leads to the value of
interior temperature $T_{\rm core} = \left( 8.98\ud{0.81}{0.98}
\right) \tee{7}\unit{K}$.  Overall, if it is assumed that the H/He
transition depth is unknown, the core temperature is in the range of
extreme values: $T_{\rm core}=\left(3.0-9.8\right)\tee{7}\unit{K}$.

\section{Conclusion}
\label{sec:conclusion}
We have performed the spectral analysis of five archived
\chandra\ observations of the qLMXBs in the GC NGC~6397.  The $\sim$
350\ksec\ of integration time available permitted to obtain high S/N
spectra and improve the radius measurement.  More specifically, the
simultaneous spectral fitting of all five observations, using an NS H
atmosphere, allowed us to provide constraints on the NS radius with
$\sim$10\% statistical uncertainty (90\% confidence).  This confirmed
the qLMXB nature of the \xray\ source.  Therefore, the measured NS
properties are $\rns=8.9\ud{0.9}{0.6}\km$ and $\kteff =
80\ud{4}{5}\eV$, for $\mns = 1.4\msun$, and assuming an NS with an
atmosphere composed of pure hydrogen.  The estimated interior
temperature lies in the range $T_{\rm core} = \left( 3.0-9.8 \right)
\tee{7}\unit{K}$.  In the 0.5--10\keV\ range, the flux corresponding
to these best-fit parameters is $F_{\rm X} = \left(
1.39\ud{0.02}{0.06} \right) \tee{-13}\cgsflux$, equivalent to a
luminosity of $L_{\rm X} = \left(1.04\ud{0.01}{0.05}\right)
\tee{32}\cgslum$ at a distance of 2.5\kpc.  The spectra did not show
evidence for a power-law component as inferred by the upper limit on
the contribution to the total flux of 3.7\%.

The results of this analysis were consistent with those of the
discovery observation (Gr01); the reported NS radius and temperature
were $\rinfty = 4.9\ud{14}{1}\km$ and $\kteff = 57$--$92\eV$.  No
optical counterpart was detected on the \hst\ observations, with a
limiting magnitude of $M_{V} > 11$ (Gr01).

No variability was observed on long timescales.  Therefore, unless an
outburst (for which there is no evidence) happened between the
observations--between 2000 and 2002, or between 2002 and 2007--we
conclude that the source remained in its quiescent stage since the
discovery observation.  It is worth noting that none of the GC qLMXBs
discovered in quiescence so far have been seen in outburst.  Moreover,
an outburst happening between the available observations would have
had an impact on the observed luminosity and intensity variability
would have been detected \citep{ushomirsky01, rutledge02c}.  The
search of short-timescale variability ($<1\unit{day}$) did not reveal
any such variability. Finally, a PDS analysis was performed and did
not demonstrate evidence of periodic variability in the frequency
range $0.0001$--$0.10\unit{Hz}$.  The lack of intensity variability on
various timescales further supports the classification of the source.

In conclusion, the qLMXB U24 in the GC NGC~6397 adds to the list of
qLMXBs suitable to place constraints on the dense matter EoSs with a
best-fit projected radius of $\rinfty = 11.9\ud{1.0}{0.8}\km$.

\acknowledgements R.E.R. is supported by an NSERC Discovery grant.
E.B. is supported by NASA/ATFP grant NNX08AG76G.  The results
presented have made use of data from archived observations available
at the High Energy Astrophysics Archive Research Center Online
Service, provided by the NASA GSFC.  Finally, the authors would like
to thank the referee for useful remarks that led to the improvement of
this article.

\bibliographystyle{apj_8}
\bibliography{biblio}

\clearpage

\clearpage

\end{document}